\documentclass[doublecol,figures]{epl2} 
\usepackage{amsmath}
\usepackage{amssymb}

\title{Simple Exclusion Processes with Local Resetting}

\author{A. Pelizzola\inst{1,2}\thanks{E-mail: \email{alessandro.pelizzola@polito.it}} \and M. Pretti\inst{1,3}\thanks{E-mail: \email{marco.pretti@polito.it}} \and M. Zamparo\inst{1} \thanks{E-mail: \email{marco.zamparo@polito.it}}}
\shortauthor{A. Pelizzola \etal}

\institute{ \inst{1} Dipartimento di Scienza Applicata e Tecnologia,
  Politecnico di Torino, Corso Duca degli Abruzzi 24, I--10129 Torino,
  Italy\\ 
\inst{2} INFN, Sezione di Torino, via Pietro Giuria 1,
  I-10125 Torino, Italy\\ 
\inst{3} Consiglio Nazionale delle Ricerche
  - Istituto dei Sistemi Complessi (CNR-ISC), Via dei Taurini 19,
  00185 Roma, Italy } 

\abstract{
We investigate the stationary state of Symmetric and Totally Asymmetric Simple Exclusion Processes with local resetting, on a one--dimensional lattice with periodic boundary conditions, using mean--field approximations, which appear to be exact in the thermodynamic limit, and kinetic Monte Carlo simulations. In both cases we find that in the thermodynamic limit the models exhibit three different regimes, depending on how the resetting rate scales with the system size. The Totally Asymmetric version of the model has a particularly rich behaviour, especially in an intermediate resetting regime where the resetting rate vanishes as the inverse of the system size, exhibiting 4 different phases, including phase separation. 
}

\begin{document}

\maketitle

Simple exclusion processes are models in which particles diffuse according to Markovian stochastic rules on a lattice, often one--dimensional, with the constraint that at most 1 particle can occupy a lattice node. They are paradigmatic models in non--equilibrium statistical physics, because of their simplicity and the rich behaviour they exhibit. In the last decades a huge amount of work has been devoted to this field (see the reviews \cite{Derrida98,Schutz,Evans2007,ZiaReview,TransportBook}), which has led to exact and numerical results for the basic models, many generalizations, powerful approximation techniques and connections with different models and research lines. 

The simplest such model is probably the one--dimensional Symmetric Simple Exclusion Process (SSEP) with periodic boundary conditions (PBCs), in which particles diffuse on a ring by hopping to adjacent empty sites, with the same hopping rate in both directions. When different hopping rates in the two directions are chosen, the model is usually called Asymmetric Simple Exclusion Process (ASEP). In the limit where hopping is allowed only in one direction we speak of the Totally Asymmetric Simple Exclusion Process (TASEP). In the case of PBCs the stationary state of these models is characterized by a uniform density of particles and converges to a product measure in the thermodynamic limit. In the case of open boundary conditions (OBCs), where at the end nodes particles can leave or enter (provided the destination node is empty) the lattice, the ASEP, as well as its limiting case TASEP, exhibit a rich stationary state and dynamical behaviour. 

Among the many generalizations and extensions of these basic models it is especially important to mention here the Totally Asymmetric Simple Exclusion Process with Langmuir Kinetics (TASEP--LK) \cite{ParmeggianiFranoschFrey03,Popkov,Evans2003,ParmeggianiFranoschFrey04}. This is a generalization of a TASEP with OBCs, in which particles can also attach to an empty node with rate $\omega_A$ or detach from an occupied one with rate $\omega_D$. Since attachment and detachment are bulk processes, the most interesting situation is obtained when the corresponding rates scale as the inverse of the system size $L$, such that the corresponding total rates $\Omega_A = L \omega_A$ and $\Omega_D = L \omega_D$ remain finite in the thermodynamic limit $L \to \infty$.

Another important direction of investigation has been opened very recently in \cite{Reuveni}, where the SSEP with local resetting (SSEP--LR) has been introduced. In this model particles diffuse as usual on a one--dimensional lattice with PBCs and the simple exclusion constraint, but they can also reset their position independently of one another, by jumping to a particular lattice node named the origin with a rate $r$. The new feature in this work was the local nature of the resetting process, which involves each particle independently of the others. In previous works on exclusion processes with resetting \cite{Pal,Nagar}, the reset was global, in the sense that the whole configuration of the system was reset to some predefined condition. As noticed in \cite{Reuveni}, local resetting
can be more challenging than its global counterpart. For instance, the approach based on renewal theory (see \cite{Majumdar} and references therein, also for a general perspective on stochastic resetting) is not applicable when resetting is applied to each particle independently. 

In \cite{Reuveni} the authors studied the SSEP--LR stationary state with $r$ of order 1 and $r \propto L$, in the (i) fixed density and (ii) fixed number of particle cases, using both mean--field (MF) approximation and Kinetic Monte Carlo (KMC) simulations. Their crucial finding is that in the thermodynamic limit the stationary density profile is independent of the resetting rate $r$, and depends only on the position (in case (i) scaled by the system size) and (i) the average density or (ii) the number of particles. In particular, the density of particles at the origin always tends to 1 as the system size $L \to \infty$. They also find a remarkable agreement between MF and KMC results. 

Local resetting has certain analogies with the Langmuir kinetics. The detachment process is similar, although in the case of local resetting the detachment rate depends on the density at the origin. On the other hand, the attachment process is completely different. Based on the analogy, one may wonder whether considering a resetting rate $r$ which vanishes in the thermodynamic limit, with a suitable scaling with $L$, could give rise to interesting scenarios. Here we try to answer this question in the case of SSEP--LR and then extend our investigation to the TASEP with local resetting (TASEP--LR).

The SSEP--LR can be defined as follows: consider a one--dimensional lattice of $L$ nodes with periodic boundary conditions. We assume $L$ even and label nodes by $l = -L/2 + 1, \cdots 0, \cdots L/2$. The node $l = 0$ will be referred to as the origin. A time--dependent occupation number variable $n_l^t$ is associated to each node, which at time $t$ can be empty ($n_l^t = 0$) or occupied by one particle ($n_l^t = 1$). The system evolution, in continuous time, is determined by 2 classes of stochastic processes: hopping and local resetting. Each particle can hop, with rate 1, to an empty neighbour node. In addition, each particle can also return to the origin, provided it is empty, with rate $r$. The total number of particles
$N = \sum_l n_l^t$
is conserved.

As a first step we shall explore the system dynamics using the MF approximation, as in \cite{Reuveni}. Introducing the local densities $\rho_l^t = \langle n_l^t \rangle$, MF corresponds to neglecting correlations and approximating $\langle n_k^t n_l^t \rangle \simeq \langle n_k^t \rangle \langle n_l^t \rangle = \rho_k^t \rho_l^t$. The time evolution of the local densities in the MF approximation is then described by the equations \cite{Reuveni}
\begin{eqnarray}
&& \hspace{-2em} \dot \rho_0^t = \rho_{1}^t - 2 \rho_0^t + \rho_{-1}^t + r (1 - \rho_0^t) \sum_{l \ne 0} \rho_l^t  \\
&&  \hspace{-2em} \dot \rho_l^t = \rho_{l+1}^t - 2 \rho_l^t + \rho_{l-1}^t - r (1 - \rho_0^t) \rho_l^t \quad (l \ne 0), 
\end{eqnarray}
with $\rho_{L/2+1} \equiv \rho_{-L/2+1}$ and $\rho_{-L/2} \equiv \rho_{L/2}$ due to PBCs. 
In the stationary state the local densities become time--independent, and we will denote them by $\rho_l$, dropping the time indices. Exploiting the symmetry $\rho_{-l} = \rho_l$ and using the definition of $N$ we obtain, for the stationary state,
\begin{eqnarray}
  && \hspace{-2em} 0 = 2 \rho_{1} - 2 \rho_0 + r (1 - \rho_0) (N - \rho_0) \\
  && \hspace{-2em} 0 = \rho_{l+1} - 2 \rho_l + \rho_{l-1} - r (1 - \rho_0) \rho_l \quad (l > 0).
  \label{eq:NESS}
\end{eqnarray}

We solve the above equations for the stationary state in a continuum limit, assuming $L \gg 1$ and $N \gg 1 \ge \rho_0$, and introducing a scaled coordinate $x = l/L \in [0,1/2]$ (the $x < 0$ portion of the profile can be obtained by symmetry). Denoting derivative with respect to $x$ by a prime we obtain
\begin{eqnarray}
  0 &=& 2 \rho'(0) + r N L (1 - \rho_0), \label{eq:boundary} \\
  0 &=& \rho''(x) - r L^2 (1 - \rho_0) \rho(x). \label{eq:continuum}
\end{eqnarray}
Defining $\lambda = \sqrt{r L^2 (1 - \rho_0)}$, assuming that $\lambda$ remains finite in the thermodynamic limit and imposing the boundary condition $\rho(0) = \rho_0$ we obtain
\begin{equation}
  \rho(x) = \rho_0 \cosh (\lambda x) - \frac{\lambda N}{2 L} \sinh(\lambda x),
  \label{eq:profile}
\end{equation}
where $\rho_0$ must be determined by imposing the condition
\begin{equation}
N = 2 L \int_0^{1/2} \rho(x) {\rm d}x = \frac{2 L \rho_0}{\lambda} \sinh \frac{\lambda}{2} - N \left( \cosh \frac{\lambda}{2} - 1 \right),
\end{equation}
which reduces to 
\begin{equation}
  \rho_0 = \frac{N}{L} \frac{\lambda}{2} \coth \frac{\lambda}{2}.
  \label{eq:rho0}
\end{equation}

Following \cite{Reuveni}, we consider separately the cases where the thermodynamic limit $L \to \infty$ is taken at fixed density $\overline{\rho} = N/L > 0$ or at vanishing density ($1 \ll N \ll L$, generalizing slightly the fixed $N$ case discussed in \cite{Reuveni}). In the fixed $\overline{\rho}$ case, the condition for $\rho_0$ eq.\ \ref{eq:rho0} becomes $\rho_0 = \overline{\rho} \frac{\lambda}{2} \coth \frac{\lambda}{2}$ and the stationary density profile becomes
\begin{equation}
  \rho(x) = \overline{\rho} \frac{\lambda/2}{\sinh(\lambda/2)} \cosh \left[ \lambda \left( \frac{1}{2} - x \right)\right].
  \label{eq:profilefixedrho}
\end{equation}
The above results suggest that the stationary density profile depends on $r$ and $L$ only through $\lambda$ (except for the prefactor $\overline{\rho}$), and hence only through the combination $r L^2$. This provides some insight into the transition from the homogeneous profile that is obtained in the purely diffusive case ($r = 0$) to the nonuniform profile, with a maximum at the origin, obtained in \cite{Reuveni} for $r$ of order 1 and $r \propto L$. Indeed, 3 different regimes can be found, depending on the behaviour of $r L^2$ (or equivalently $r N^2$) in the thermodynamic limit $L \to \infty$. (i) Small resetting: if $r$ tends to 0 faster than $L^{-2}$ (or equivalently $N^{-2}$), then $\lambda \to 0$, $\rho_0 = \overline{\rho}$ and $\rho(x) = \overline{\rho}$, the purely diffusive case. (ii) Large resetting: if $r L^2 \to \infty$, which includes the cases studied in \cite{Reuveni}, $\rho_0$ tends to $1$ (more precisely $\rho_0 = 1 - \lambda/(r L^2)$ and $\lambda$ stays finite) and the profile has a maximum at the origin. In this regime local resetting dominates over diffusion. (iii) Intermediate resetting: if $r$ tends to $0$ as $L^{-2}$ (equivalently $N^{-2}$), $\lambda$ tends to a positive constant, $\rho_0 \in (\overline{\rho},1)$ and again the profile has a maximum at the origin. In this regime, illustrated in Fig.\ \ref{fig:ProfilesFixedDensity} in the case $\overline{\rho} = 0.2, r L^2 = 100$, local resetting and diffusion are in a balanced competition. The MF stationary density profile is plotted together with profiles from KMC simulations for 3 different lattice sizes. KMC simulations are carried out using Gillespie algorithm, running time is $10^6$ and averages are taken in the stationary state, for $t \in (10^5, 10^6)$. The collapse of KMC data is remarkable, as well as the agreement with the MF results, the profiles are almost indistinguishable on the drawing scale. In order to further check the accuracy of the MF approximation we also plotted the nearest--neighbour (NN) correlations $c(x=l/L) = \langle n_l n_{l+1} \rangle - \langle n_l \rangle \langle n_{l+1} \rangle$, which seem to vanish. This, together with the MF accuracy found in \cite{Reuveni} in regime (ii), and with the exactness of MF for the purely diffusive case, suggests that MF may be exact for this model in the thermodynamic limit. 

\begin{figure}
\onefigure[width=0.45\textwidth]{ProfilesFixedDensity.eps}
\caption{Stationary density profiles (top lines) and NN correlations (bottom line) for the SSEP--LR, $\overline{\rho} = 0.2, r L^2 = 100$. MF density profile: black. KMC density profile: red ($L = 500$), blue ($L = 1000$), green ($L = 2000$). KMC NN correlations: red ($L = 500$).}
\label{fig:ProfilesFixedDensity}
\end{figure}

In the vanishing density case, the condition for $\rho_0$ eq.\ \ref{eq:rho0} becomes $\rho_0 = \sqrt{r N^2 (1 - \rho_0)}/2$, 
which yields
$\rho_0 = \frac{1}{8} r N^2 ( \sqrt{1 + 16/(r N^2)} - 1 )$.
The stationary density profile eq.\ \ref{eq:profile} is more conveniently rewritten as 
\begin{equation}
  \rho(y) = \rho_0 \cosh (\mu y) - \frac{\mu}{2} \sinh(\mu y),
\end{equation}
with new scaled variables $\mu = \sqrt{r N^2 (1 - \rho_0)}$ and $y = l/N$.

\begin{figure}
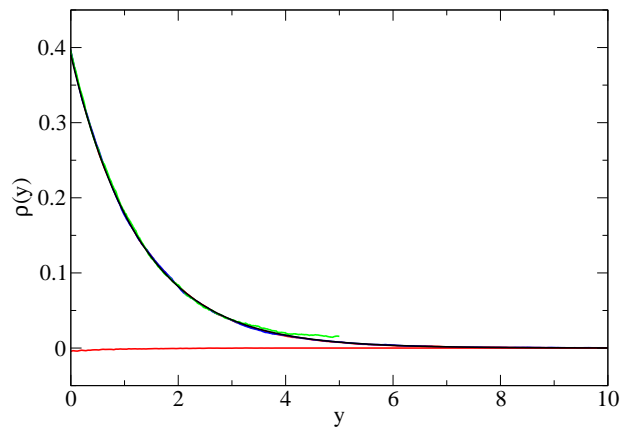

\onefigure[width=0.45\textwidth]{ProfilesVanishingDensity.eps}
\caption{Stationary density profiles (top lines) and NN correlations (bottom line) for the SSEP--LR, $L = 1000, r N^2 = 1$. MF density profile: black. KMC density profile: red ($N = 25$), blue ($N = 50$), green ($N = 100$). KMC NN correlations: red ($N = 25$).}
\label{fig:ProfilesVanishingDensity}
\end{figure}

As $L \to \infty$, the behaviour of $\rho_0$ is now determined by $r N^2$, and again we can find 3 different regimes, similarly to the finite density case. (i) If $r N^2 \to 0$ ($r$ tends to 0 faster than $N^{-2}$), then $\rho_0 \to 0$, purely diffusive case. (ii) If $r N^2 \to \infty$, local resetting dominates over diffusion, $\rho_0$ tends to $1$ (more precisely $\rho_0 \simeq 1 - 4/(r N^2)$) and the profile has a maximum at the origin and covers a finite portion of the lattice (as shown in \cite{Reuveni}). (iii) Finally, if $r N^2$ tends to a positive constant ($r$ tends to $0$ as $N^{-2}$), local resetting and diffusion are in balanced competition, $\rho_0 \in (0,1)$, the profile has again a maximum at the origin and covers a finite portion of the lattice. This regime is illustrated in Fig.\ \ref{fig:ProfilesVanishingDensity}. The agreement between KMC and MF, and the collapse of the KMC data are again remarkable, except for some finite size effect in the case $N = 100$. 

We now turn our attention to the asymmetric version of the model, in which the rightward and leftward hopping rates are different. In particular, we focus on the TASEP, where only rightward (from $l$ to $l+1$) hopping is allowed, with rate $1$. In the MF approximation, the time evolution equations for the local densities in TASEP--LR are
\begin{eqnarray}
&& \hspace{-3em} \dot \rho_0^t = - \rho_0^t (1 - \rho_1^t) + \rho_{L-1}^t (1 - \rho_0^t) + r (1 - \rho_0^t) \sum_{l=1}^{L-1} \rho_l^t \\
&& \hspace{-3em}  \dot \rho_l^t = - \rho_l^t (1 - \rho_{l+1}^t) + \rho_{l-1}^t (1 - \rho_l^t) - r (1 - \rho_0^t) \rho_l^t  \nonumber \\
  && \hspace{-3em} (l = 1, \ldots L-1).  
\end{eqnarray}
Now it is more convenient to let $l$ take values from 0 to $L-1$, since the system is no longer symmetric with respect to the origin. Due to the periodic boundary conditions, $\rho_L \equiv \rho_0$.
The stationary state equations are therefore
\begin{eqnarray}
&& \hspace{-3em} 0 = - \rho_0 (1 - \rho_1) + (1 - \rho_0) [ \rho_{L-1} + r (N - \rho_0)] \label{eq:TASEP-NESS0} \\
&& \hspace{-3em} 0 = - \rho_l (1 - \rho_{l+1}) + \rho_{l-1} (1 - \rho_l) - r (1 - \rho_0) \rho_l \nonumber \\
&& \hspace{-3em} (l = 1, \ldots L-1). 
\label{eq:TASEP-NESS}
\end{eqnarray}
It can be verified that eq.\ \ref{eq:TASEP-NESS} is equivalent to the MF stationary state equation for a TASEP--LK with OBCs \cite{ParmeggianiFranoschFrey04} and $\rho_0$--dependent parameters: injection rate $\rho_0$ at $l = 1$, extraction rate $1 - \rho_0$ at $l = L - 1$, total attachment rate $\Omega_A = 0$ and total detachment rate $\Omega_D = r L (1 -\rho_0)$. The resetting current into the origin generates a discontinuity, an effect analogous to opening the lattice. 

Taking the continuum limit of eq.\ \ref{eq:TASEP-NESS} as in the SSEP case (now with $x \in [0,1)$) we obtain
\begin{equation}
\frac{{\rm d}}{{\rm d}x} \left[ \rho(x) \left( 1 - \rho(x) \right) \right] = - r L (1 - \rho_0) \rho(x), 
\end{equation}
more conveniently rewritten in terms of the function $f(x) = F(\rho(x))$, where $F(\rho) = \rho e^{- 2 \rho}$, as
\begin{equation}
f'(x) = - r L (1 - \rho_0) f(x).
\end{equation}

A first consequence of this result is that the behaviour of the TASEP--LR will exhibit different regimes depending on the value of $r L$. If $r L$ tends to $0$ in the thermodynamic limit then $\rho(x) \equiv \rho_0$ and we are in a small resetting, purely diffusive regime. 
If $\lambda = r L (1 - \rho_0)$ stays finite in the thermodynamic limit $L \to \infty$ 
we obtain
\begin{equation}
  f(x) = 
  {\rm const} \cdot e^{- \lambda x}.
  \label{eq:LambertProfile}
\end{equation}
Moreover, eq.\ \ref{eq:TASEP-NESS0} tells us that the current has a finite discontinuity at the origin. Summing it with eq.\ \ref{eq:TASEP-NESS} for $l = 1$ and $l = L - 1$ and neglecting terms of order 1 we obtain 
\begin{equation}
J_* \equiv J_+ - J_- = r N (1 - \rho_0) = \overline{\rho} \lambda.
  \label{eq:Currents}
\end{equation}
Here $J_+ = \rho_1 (1 - \rho_2)$ (respectively $J_- = \rho_{L-2} (1 - \rho_{L-1})$) denotes the current out of the origin, to the right (resp.\ into the origin, from the left), and $J_*$ is the current into the origin due to resetting (we have neglected $\rho_0$ with respect to $N$). In the continuum limit we obtain
\begin{equation}
J_\pm = \rho_\pm (1-\rho_\pm),
\label{eq:Jpm}
\end{equation}
where $\rho_+ = \lim_{x \to 0^+} \rho(x)$ and $\rho_- = \lim_{x \to 1^-} \rho(x)$. The average density can then be written, using eq.\ \ref{eq:Currents} as
\begin{equation}
\overline{\rho} = \frac
{\left( \rho_- - \frac{1}{2} \right)^2 
-\left( \rho_+ - \frac{1}{2} \right)^2}
{rL (1-\rho_0)}.
\label{eq:averhovsrho0}
\end{equation}

Before turning these results into results for the density profile, a few remarks are in order. The function $f = F(\rho) = \rho e^{-2 \rho}$ is increasing for $\rho \in (0,1/2)$, has a maximum at $\rho = 1/2$ and then decreases for $\rho \in (1/2,1)$. For $f \in (1/e^2,1/(2e))$, its inverse is not single--valued. Indeed, it can be written in terms of a Lambert $W$ function as $\rho = - W(- 2 f)/2$. $W$ has 2 real branches, the so--called principal branch $W_0(z)$ and a second branch $W_{-1}(z) \le W_0(z)$, with equality for $z = -1/e$. As a consequence, we have to consider two possible solutions for our density profile, a low--density (LD) one $\rho_{LD}(x) = - W_0(- 2 f(x))/2 \in (0,1/2)$ for $f(x) \in (0, 1/(2 e))$, and a high--density (HD) one $\rho_{HD}(x) = - W_{-1}(- 2 f(x))/2 \in (1/2,1)$ for $f(x) \in (1/e^2, 1/(2 e))$. Keeping this in mind, we now proceed to a discussion of the stationary state.

\begin{figure}
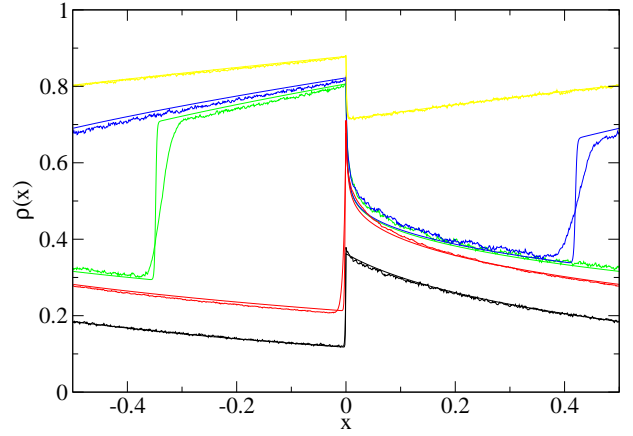

\onefigure[width=0.45\textwidth]{TASEP2branches_L1000_r1e-3.eps}
\caption{Stationary density profiles for the TASEP--LR in the intermediate resetting regime for $L = 10^3$, $r = 10^{-3}$ and several values of $\overline{\rho}$. Smooth lines: MF approximation. Noisy lines: KMC. $\overline{\rho} = 0.2$ (black), 0.3 (red), 0.5 (green), 0.6 (blue) and 0.8 (yellow). The $x$ range has been shifted to $(-1/2,1/2)$ for clarity by taking advantage of PBCs.}
\label{fig:TASEP-rL1}
\end{figure}

We consider a fixed, finite value of $rL$, corresponding to an intermediate resetting regime (we will discuss the large resetting regime later) and an average density $\overline{\rho} \in (0,1)$. Representative density profiles from MF approximation and KMC simulations (running time is $10^6$ and averages are taken in the stationary state, for $t \in (10^5, 10^6)$) are shown in Fig.\ \ref{fig:TASEP-rL1} in the case $r L = 1$. 

\begin{figure}
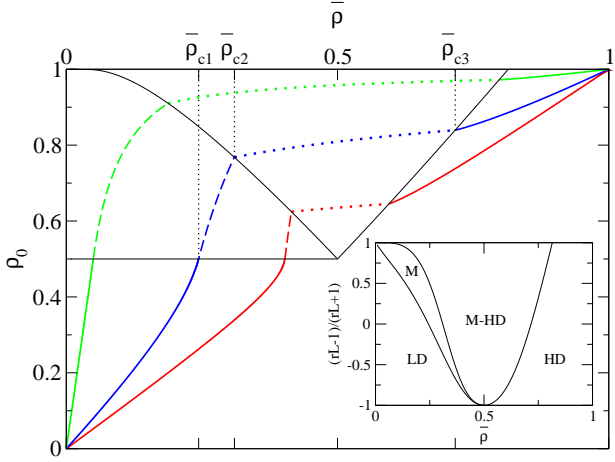

\onefigure[width=0.45\textwidth]{PhaseDiagrams.eps}
\caption{Local density $\rho_0$ at the origin in the TASEP--LR as a function of the average density $\overline{\rho}$ for $rL = 0.1$ (red), 1 (blue) and 10 (green). Full lines: LD and HD phases. Dashed lines: M phases. Dotted lines: M--HD phase separation. Black lines denote phase transitions. Inset: phase diagram in the $(\overline{\rho},(rL-1)/(rL+1))$ plane.}
\label{fig:gap}
\end{figure}

As $\overline{\rho}$ grows, the local density $\rho_0$ at the origin also grows from 0 to 1, as illustrated in Fig.\ \ref{fig:gap}, and the parameters of the equivalent TASEP--LK model vary, describing a line in its phase diagram. The stationary state goes through 4 different phases, separated by 3 transitions at average density $\overline{\rho}_{c1-3}$. These transitions are shown in Fig.\ \ref{fig:gap} with black lines, marking explicitly the case $rL = 1$. The phase diagram in terms of the control parameters $\overline{\rho}$ and $rL$ is reported in the inset, where the vertical coordinate has been chosen as $(rL-1)/(rL+1)$ for convenience. 

We now describe the 4 phases which are encountered for increasing average density. In particular, for each phase, we shall give $\rho_\pm$ (thence $\overline{\rho}$, by eq.\ \ref{eq:averhovsrho0}) as a function of $\rho_0$. Notice that $\rho_\pm$, except in the case of phase separation, are related through eq.\ \ref{eq:LambertProfile} by the condition
\begin{equation}
F(\rho_-) = e^{-rL(1-\rho_0)} F(\rho_+) = e^{-\lambda} F(\rho_+). 
\label{eq:Fpure}
\end{equation}

For $\overline{\rho} < \overline{\rho}_{c1}$ we find a LD solution, with $\rho_{LD}'(x) < 0$, 
\begin{equation}
\rho_+ = \rho_0, \qquad  
\rho_- 
= - {\textstyle \frac{1}{2}} W_0 \left( -2 F(\rho_0) e^{-rL(1-\rho_0)} \right).
\label{eq:rhopmLD}
\end{equation}
The density discontinuity at the origin (more precisely, immediately on the left of the origin) corresponds to the discontinuity in the current. $\rho_0$ is an increasing function of $\overline{\rho}$, implicitly given by eqs.\ \ref{eq:averhovsrho0} and \ref{eq:rhopmLD} and shown in Fig.\ \ref{fig:gap}, which reaches the limiting value $1/2$ at the transition value $\overline{\rho}_{c1}$.

\begin{figure}
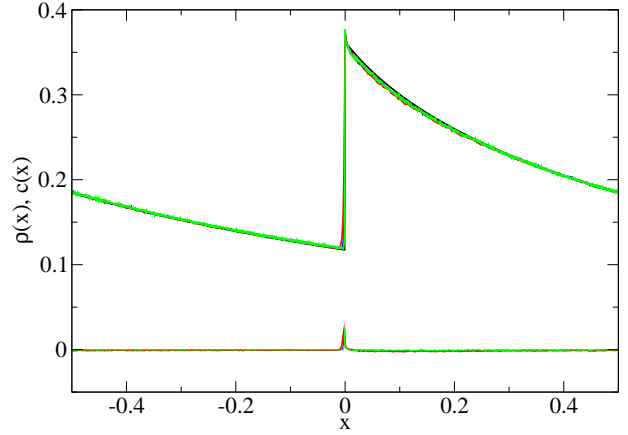

\onefigure[width=0.45\textwidth]{TASEP_density02_rL1.eps}
\caption{Stationary density profiles and NN correlations for the TASEP--LR, $\overline{\rho} = 0.2, r L = 1$. MF density profile: black. KMC density profile (top lines) and NN correlations (bottom lines): red ($L = 500$), blue ($L = 1000$), green ($L = 2000$).}
\label{fig:TASEP-LD}
\end{figure}

Our results suggest that the stationary density profiles depend on the parameters $r$ and $L$ only through their combination $rL$. It is therefore interesting to check this scaling behaviour as $L \to \infty$. In Fig.\ \ref{fig:TASEP-LD} we plot the MF stationary density profile in the case $r L = 1$, $\overline{\rho} = 0.2$ together with profiles and NN correlations from 
KMC simulations for 3 different lattice sizes. As in the SSEP--LR case, the collapse of KMC data is remarkable, as well as the agreement with the MF results. The NN correlations seem to vanish everywhere except close to the origin. The largest of these correlations is always $c(x=-1/L) = \langle n_{L-1} n_0 \rangle - \langle n_{L-1} \rangle \langle n_0 \rangle$. This result suggests that MF may be exact except in a small (vanishing, on the scale of the lattice size, in the thermodynamic limit) region around the discontinuity. 

\begin{figure}
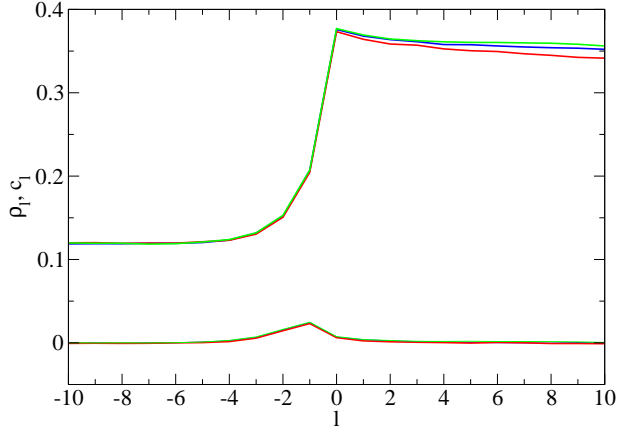

\onefigure[width=0.45\textwidth]{TASEP_density02_rL1NotScaled.eps}
\caption{KMC results in Fig.\ \protect\ref{fig:TASEP-LD} as functions of the node index $l$.}
\label{fig:TASEP-LD-NotScaled}
\end{figure}

In order to better understand the region close to the origin, the KMC results are plotted in Fig.\ \ref{fig:TASEP-LD-NotScaled} as a function of the node index $l$ (instead of the scaled coordinate $x = l/L$). On the left of the origin, in a microscopic (with respect to the lattice size $L$) region, whose width is a few lattice nodes, density profiles and correlations turn out to be well--defined functions of the scaled resetting rate $r L$ and the position $l$ (not scaled), a behaviour which resembles that of a boundary layer in pure TASEP and in TASEP--LK. 

\begin{figure}
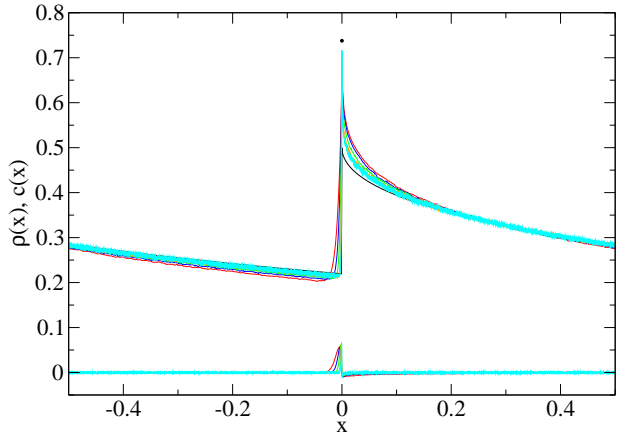

\onefigure[width=0.45\textwidth]{TASEPdensity03rL1.eps}
\caption{Stationary density profiles and NN correlations for the TASEP--LR, $\overline{\rho} = 0.3, r L = 1$. MF density profile: black. KMC density profile (top lines) and NN correlations (bottom lines): red ($L = 500$), blue ($L = 1000$), green ($L = 2000$), orange ($L = 5000$), cyan ($L = 10000$).}
\label{fig:TASEP-MC}
\end{figure}

For $\overline{\rho} \in (\overline{\rho}_{c1},\overline{\rho}_{c2})$ (e.g.\ $\overline{\rho} = 0.3$ in Fig.\ \ref{fig:TASEP-rL1}) a different solution is found, still of the LD type for $x \in (0,1)$, but now with 
\begin{equation}
\rho_+ = {\textstyle \frac{1}{2}} < \rho_0, \ 
\rho_- 
= - {\textstyle \frac{1}{2}} W_0 \left( - 2 F(\rho_+) e^{-rL(1-\rho_0)} \right). 
\label{eq:rhopmMC}
\end{equation}
Physically, $\rho_0$ determines the resetting current and the shape of the profile, and the current $J_+ = 1/4$ is maximal. For this reason, this phase is called maximal current (M) in TASEP--LK, and we will follow this convention. In a finite system, a microscopic boundary layer forms, joining the local density at the origin with the bulk profile. This is shown in Fig.\ \ref{fig:TASEP-MC} for $\overline{\rho} = 0.3$, where $\rho_0 \simeq 0.738$ (MF, continuum limit) is represented by an isolated point. The scaling of KMC results, and the agreement with MF, is again very good, but the approach to the thermodynamic limit is now much slower than in the pure LD phase. 

For $\overline{\rho} \in (\overline{\rho}_{c2},\overline{\rho}_{c3})$ (e.g.\ $\overline{\rho} = 0.5$ or 0.6 in Fig.\ \ref{fig:TASEP-rL1}) the system exhibits phase separation into 2 pure phases: a M phase on the right of the origin, and a HD phase (with $\rho_{HD}'(x) > 0$) on the left. 
The mean--field stationary density profile is a piecewise combination of a M portion for ${x \in (0,x_s)}$ and a HD one for ${x \in (x_s, 1)}$, with 2 domain walls (DWs). 
One DW is (as usual) at the origin, where the density jumps downward from $\rho_- = \rho_0$ to $\rho_+ = 1/2$. 
Eq.\ \ref{eq:averhovsrho0} thus reads $\overline{\rho} = (\rho_0 - 1/2)^2 / \left[ rL (1-\rho_0) \right]$, shown in Fig.\ \ref{fig:gap}  with a dotted line. 
The other DW is at ${x = x_s}$, where the density jumps upward from $\rho_s < 1/2$ to $1-\rho_s > 1/2$, such that the current is continuous, according to eq.\ \ref{eq:TASEP-NESS}. 
Quantitatively, we can write
\begin{equation}
\rho(x) 
= 
\begin{cases}
- {\textstyle \frac{1}{2}} W_0 \left( -2 F(\rho_+) e^{-\lambda x} \right) 
&  x \in (0,x_s)
\\
- {\textstyle \frac{1}{2}} W_{-1} \left( -2 F(\rho_-) e^{\lambda (1-x)} \right) 
& x \in (x_s,1)
\end{cases}
\end{equation}
The position $x_s$ of the extra DW and the density $\rho_s$ can then be obtained from the conditions ${\lim_{x \to x_s^-} \rho(x) = \rho_s}$ and ${\lim_{x \to x_s^+} \rho(x) = 1-\rho_s}$ that, according to eq.~25, read
\begin{align}
F(\rho_s) & = F({\textstyle \frac{1}{2}}) \, e^{-rL(1-\rho_0) x_s}
\\
F(1-\rho_s) & = F(\rho_0) \, e^{rL(1-\rho_0) (1-x_s)}
\end{align}
The position $x_s$ of this DW decreases with $\overline{\rho}$, and the 2 transition values $\overline{\rho}_{c2}$ and $\overline{\rho}_{c3}$ can be obtained by imposing $x_s = 1$ and $x_s = 0$ respectively.

\begin{figure}
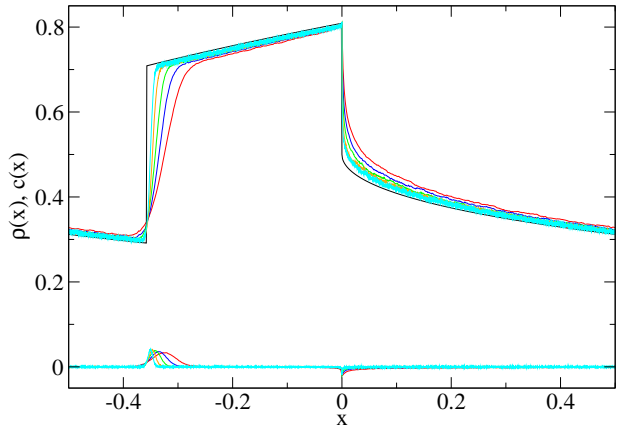

\onefigure[width=0.45\textwidth]{TASEPdensity05rL1.eps}
\caption{Stationary density profiles and NN correlations for the TASEP--LR, $\overline{\rho} = 0.5, r L = 1$. MF density profile: black. KMC density profile (top lines) and NN correlations (bottom lines): red ($L = 500$), blue ($L = 1000$), green ($L = 2000$), orange ($L = 5000$), cyan ($L = 10000$).}
\label{fig:TASEP-PhaseSep-Intermediate}
\end{figure}

The scaling behaviour of this phase separation phenomenon is shown in Fig.\ \ref{fig:TASEP-PhaseSep-Intermediate} for $\overline{\rho} = 0.5$. The KMC density profile seems to tend to the MF one as the system size $L$ grows, although the approach is very slow, apparently slower than in the M phase. In particular, the DWs become steeper as $L$ grows (numerically, we observe that the DW width is compatible with the TASEP--LK $L^{-1/2}$ scaling \cite{ParmeggianiFranoschFrey03}, though a more refined analysis is certainly worth), and NN correlations vanish everywhere except at DWs.

Finally, for $\overline{\rho} > \overline{\rho}_{c3}$ (e.g.\ $\overline{\rho} = 0.8$ in Fig.\ \ref{fig:TASEP-rL1}) we find a pure HD solution, with
\begin{equation}
\rho_- = \rho_0,
\qquad 
\rho_+ 
= - {\textstyle \frac{1}{2}} W_{-1} \left( -2 F(\rho_0) e^{rL(1-\rho_0)} \right).
\end{equation}
The density discontinuity is now immediately on the right of the origin. Results for this phase exhibit scaling behaviours and boundary layers which are qualitatively similar to the LD phase, so we omit their detailed discussion.

Regarding the transition lines shown in Fig.\ \ref{fig:gap}, we observe that, at the phase transitions, $\overline{\rho}, \rho_0$ and $r L$ can be written as functions of $J_*$. 
Indeed, in pure phases we can solve eq.\ \ref{eq:Fpure} for $\lambda$, yielding $\lambda = \ln (\rho_+/\rho_-) - 2(\rho_+ - \rho_-)$. This allows to write $\overline{\rho} = J_*/\lambda$ and $r L = \lambda/(1 - \rho_0)$ as functions of $\rho_0$ and $\rho_\pm$. 
From eq.\ \ref{eq:Currents} and eq.\ \ref{eq:Jpm} we also find $\rho_- = \frac{1}{2} \pm \sqrt{J_* + \left( \rho_+ - {\textstyle \frac{1}{2}} \right)^2}$, where the $+$ (respectively $-$) sign applies to the HD (resp.\ LD and M) phase. The 3 transition lines can then be obtained by plugging the appropriate conditions in the above equations and by varying $J_* \in (0,1/4)$.

The continuous LD--M transition, occurring at $\overline{\rho} = \overline{\rho}_{c1}$, is characterized by $\rho_+ = \rho_0 = 1/2$, from which $\rho_- =  \frac{1}{2} - \sqrt{J_*}$. The transition between the M phase and the M--HD phase separation, at $\overline{\rho} = \overline{\rho}_{c2}$, is characterized by $\rho_+ = 1/2$ and $\rho_- = 1 - \rho_0$ (corresponding to $x_s = 1$), hence $\rho_- = \frac{1}{2} - \sqrt{J_*}$ and $\rho_0 = \frac{1}{2} + \sqrt{J_*}$. Finally, the transition between the M--HD phase separation and the HD phase, at $\overline{\rho} = \overline{\rho}_{c3}$, is characterized by $\rho_+ = 1/2$ and $\rho_- = \rho_0$ (corresponding to $x_s = 0$), which yields $\rho_- = \rho_0 = \frac{1}{2} + \sqrt{J_*}$. 

\begin{figure}
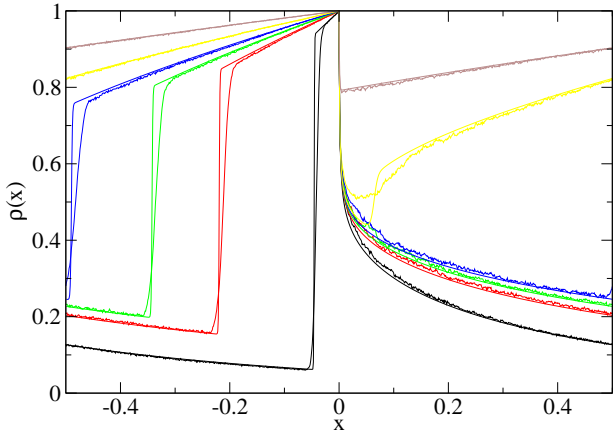

\onefigure[width=0.45\textwidth]{TASEP2branches_L1000_r1.eps}
\caption{Same as Fig.\ \protect\ref{fig:TASEP-rL1} for $r = 1$. $\overline{\rho} = 0.2$ (black), 0.4 (red), 0.5 (green), 0.6 (blue), 0.8 (yellow) and 0.9 (brown).}
\label{fig:TASEP-rL1000}
\end{figure}

The large resetting regime $r L \to \infty$ is illustrated in Fig.\ \ref{fig:TASEP-rL1000} in the case $rL = 10^3$. $\rho_0$ is always practically 1 (exactly as $rL \to \infty$), a pure HD profile is observed at large average density (precisely, as $rL \to \infty$, for $\overline{\rho} > 1/(4(1 - \ln 2))$, a value which corresponds to $\lambda = 1 - \ln 2$) and a pure LD or M profile would be observed only for a very small average density, which tends to 0 as $rL \to \infty$ (see Fig.\ \ref{fig:gap}). The profiles are qualitatively similar to those in the intermediate regime, except for the value of $\rho_0$. 

The phase separation phenomenon in the large resetting regime is qualitatively similar to the one we have just discussed, the main difference being that in this case $\rho_0 = 1$.

Finally, in the case of vanishing density ($1 \ll N \ll L$), the stationary state is determined by the parameter $r N$, with 3 regimes which are analogous to the SSEP case. In the thermodynamic limit, if $r N \to 0$ we have the purely diffusive case, with $\rho_0 \to 0$, otherwise we have an asymmetric density profile, which covers a finite portion of the lattice on the right of the origin, whose density $\rho_0 \in (0,1)$ if $r N$ tends to a finite value, while $\rho_0 \to 1$ if $r N \to \infty$. In both cases, the proper coordinate is $y = l/N$. 


Summarizing, we have shown that the stationary state of SSEP--LR and TASEP--LR in the thermodynamic limit depends crucially on how the resetting rate $r$ scales with the system size $L$. In SSEP--LR with finite density we find a small resetting, purely diffusive regime if $rL^2 \to 0$, an intermediate resetting regime if $r L^2$ tends to a positive constant, and the large resetting regime investigated in \cite{Reuveni} if $r L^2 \to \infty$. In the vanishing density case similar considerations apply, with the driving parameter $r N^2$. In TASEP--LR we have a similar picture in terms of the parameter $r L$ ($r N$ in case of vanishing density) and we suggest an analogy between TASEP--LR with PBCs and TASEP--LK with OBCs. The intermediate resetting regime of TASEP--LR is especially interesting, since the stationary state exhibits 4 different phases (3 pure phases and a phase separation), separated by 3 phase transitions. In all cases the agreement between MF and KMC, as the system size grows, is remarkable. 

These results suggest several possible lines of further investigation, and work is in progress along at least some of these lines. A first question is how an ASEP with local resetting would bridge the SSEP--LR and TASEP--LR results, in particular it would be interesting to understand whether an arbitrarily small asymmetry is sufficient to induce the rich behaviour that we have observed in TASEP--LR. Another natural step forward would be to introduce local resetting in TASEP (or, more generally, ASEP) with OBCs, and in model with additional interactions \cite{Katz1984,Antal2000,Dierl2012,Dierl2013}. Regarding methods, based on the agreement between MF and KMC found in \cite{Reuveni} and in our work, it would be interesting to rigorously assess whether, and to what extent, MF results can become exact in the thermodynamic limit. It would also be worth investigating the relaxation towards the stationary state, in which dynamical transitions without a static counterpart have been found in (T)ASEP with OBCs \cite{deGierEssler05,deGierEssler08,ProemeBlytheEvans11}, also (at least with approximate methods) in the presence of Langmuir kinetics \cite{Botto2019,Botto2020} or additional interactions between particles \cite{Botto2018,Puccioni}. Finally, we hope that these theoretical results can stimulate progress in the experimental studies, which as far as we know have been so far limited to resetting in single--particle systems \cite{Roichman,Ciliberto}.


\end{document}